\def\gs{$\Gamma_{\rm soft}$}
\def\civ{{\sc{Civ}}$\lambda$1549\/}
\def\civbc{{\sc{Civ}$_{BC}$}$\lambda$1549\/}
\def\ltsima{$\; \buildrel < \over \sim \;$}
\def\simlt{\lower.5ex\hbox{\ltsima}}
\def\hb{{\sc{H}}$\beta$\/}
\def\hbbc{{\sc{H}}$\beta_{\rm BC}$\/}

\documentstyle[aaspp4]{article}
\received{} \accepted{}
\def\rfe{R$_{\rm FeII}$}
\def\feiiq{\rm Fe{\sc ii }$\lambda$4570\/}
\def\REF{\par\noindent\hangindent 20pt}

\def\ltsima{$\; \buildrel < \over \sim \;$}
\def\simlt{\lower.5ex\hbox{\ltsima}}            % < over MMM
\def\gtsima{$\; \buildrel > \over \sim \;$}

\def\simgt{\lower.5ex\hbox{\gtsima}}            % > over MMM

\def\civ{{\sc{Civ}}$\lambda$1549\/}

\def\civbc{{\sc{Civ}}$\lambda$1549$_{\rm BC}$\/}

\def\cm3{cm$^{-3}$\/}
\def\hb{{\sc{H}}$\beta$\/}

\def\hbbc{{\sc{H}}$\beta_{\rm BC}$\/}

\def\o4363{{\sc{[Oiii]}}$\lambda$4363\/}

\def\feii{{\sc{Feii}}$_{\rm opt}$\/}
\def\fe{{\sc{Fe}}\/}

\def\fe76087{{\sc [Fe vii]}$\lambda$6087\/}
\def\oiii{{\sc [Oiii]}$\lambda$5007}

\def\kms{km~s$^{-1}$}

\begin{document}
\title{\sc Eigenvector 1: An Optimal Correlation Space for Active Galactic Nuclei}

\author{J. W. Sulentic\altaffilmark{1}, T. Zwitter\altaffilmark{2}, P. Marziani\altaffilmark{3}
and D. Dultzin-Hacyan\altaffilmark{4}}

\altaffiltext{1}{Department of  Physics and Astronomy, University
of Alabama, Tuscaloosa, AL 35487; giacomo@merlot.astr.ua.edu}

\altaffiltext{2}{Department of Physics, University of Ljubljana,
Jadranska 19, 1000 Ljubljana, Slovenia; tomaz.zwitter@uni-lj.si}

\altaffiltext{3}{Osservatorio Astronomico di Padova, vicolo
dell'Osservatorio 5, I-35122 Padova, Italy; marziani@pd.astro.it}

\altaffiltext{4}{Instituto de Astronomia, UNAM, Mexico, DF 04510,
Mexico; deborah@astroscu.unam.mx}


\begin{abstract}
We identify a correlation space involving optical and UV emission
line parameters as well as the soft X-ray spectral index that
provides optimal discrimination between all principal classes of
active galactic nuclei. Most of the sources in our three high
quality data samples show a strong intercorrelation with narrow
line Seyfert 1 (NLSy1) galaxies and steep spectrum radio galaxies
occupying opposite extrema in the space.  NLSy1 sources show a
clear continuity with broader line sources indicating that they
are not a disjoint class of AGN as is sometimes suggested. We
interpret the principal intercorrelation in the parameter space
as being driven by AGN luminosity to black hole mass ratio (L/M
$\propto$ Eddington ratio).  Source orientation no doubt also
plays an important role but it is not yet clear whether FWHM \hb\
or \civ\ line shift is the better indicator.  We tentatively
identify two RQ populations: an almost pure RQ population A with
FWHM$\leq$4000 and population B which occupies the same parameter
domain as the flat spectrum RL sources. A possible interpretation
sees Pop. A/NLSy1 as lower mass/high accretion rate sources and
Pop. B/RL sources the opposite.

\end{abstract}

\section{Introduction}

The search for correlations among observational parameters that
describe AGN has been developing rapidly in the past few years.
At the same time there has been a theoretical countercurrent that
views all quasar spectra as remarkably similar. This erroneous
view is due in part to the low resolution and s/n of much
available spectral data. Blurred quasar spectra do look
remarkably similar but data with s/n$>$20 in the continuum and
resolution $\sim$5\AA\ show striking differences from which a
pattern is beginning to emerge.  Some of the most promising
correlations arose when high s/n optical spectra were published
for the lower redshift (z$\simlt$0.5) sources in the Bright
Quasar Survey (Boroson \& Green 1992:  BG92). A principal
component analysis of the BG92 correlation matrix showed
``eigenvector 1'' correlations involving the widths and strengths
of [OIII]$\lambda$5007, as well as broad \hb\ and \feii\ emission
lines. More recently measures of X-ray luminosity, particularly
the soft X-ray photon index, have emerged as a related part of
the eigenvector 1 correlations (Wang, Brinkmann \& Bergeron 1996).

We report a study of correlations involving the best available
data samples for low redshift (z$\simlt$1.0) Active Galactic
Nuclei (AGN). We find a correlation space where we are able to
discriminate between all of the major forms of AGN phenomenology
(e.g.  broad and narrow Line Seyfert 1 galaxies, regular and
broad absorption line (BAL) quasars, steep and flat spectrum
broad line radio galaxies).  We refer to it as the Eigenvector 1
(hereafter E1) space reflecting its partial roots in the BG92
study. The three principal ``orthogonal'' correlates involve
measures of:  1) low ionization broad line width - full width at
half maximum of \hb\ (FWHM \hbbc), 2) ratio of line strengths -
the continuum normalized ratio of the optical \feii\ and broad
\hb\ emission line strengths (\rfe = W(\feiiq\ blend)/W(\hbbc)
and 3) X-ray continuum strength - the soft X-ray photon index
(\gs). Parts of this correlation space have been discussed for
almost ten years (e.g. Boroson \& Green 1992 (PG92); Boller et
al. 1996; Marziani et al.  1996 (MS96); Wang et al., 1996; Laor
et al.  1997) but the pieces have not previously been united in
this way. We present the basic E1 phenomenology in this paper. In
a following paper we will consider the principal physical drivers
of the E1 correlations.

\section{The E1 Correlation Space}

\begin{figure}
\figurenum{1} \plotone{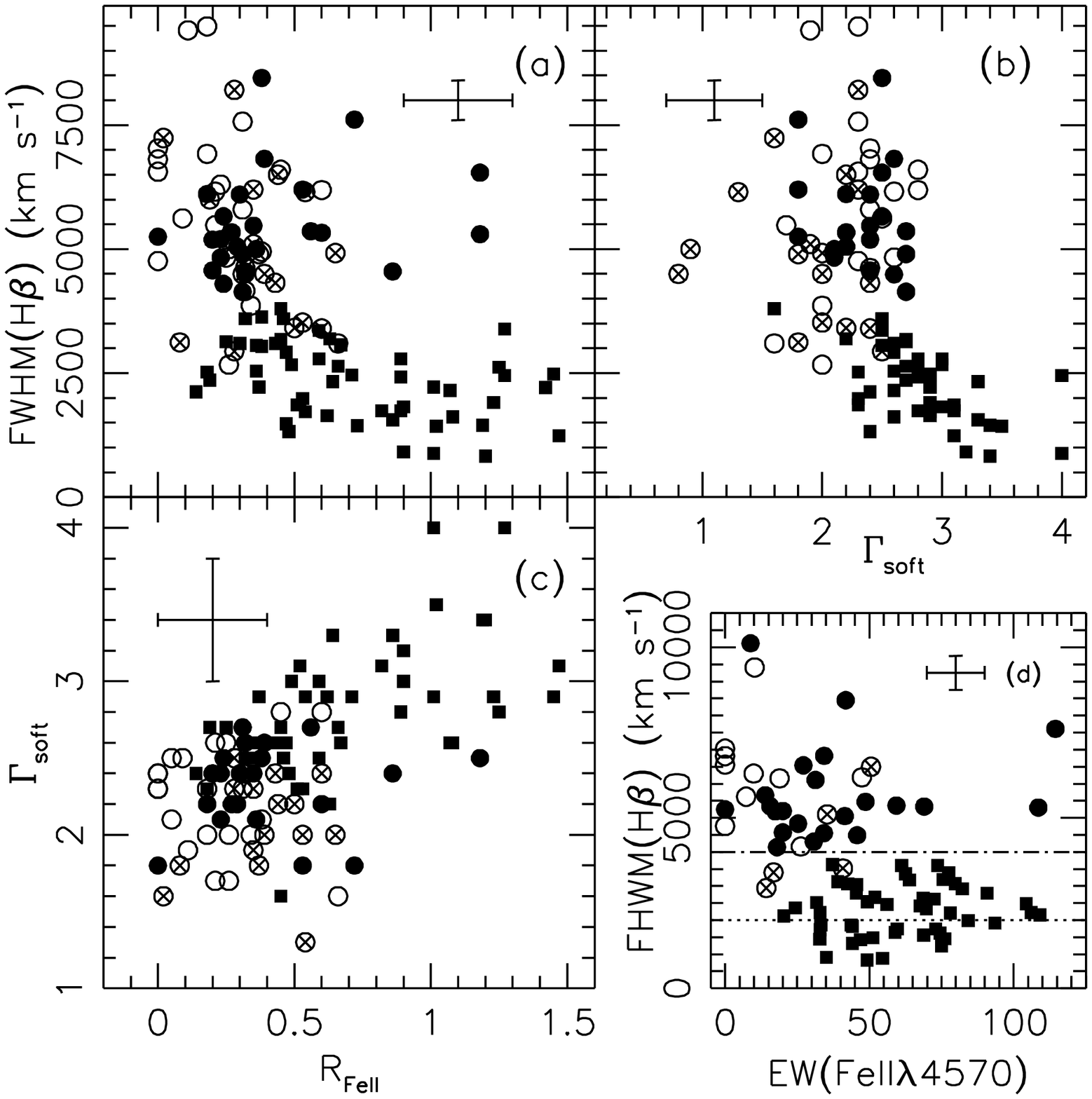} \caption[1]{Panels a,b,c: The
three principal correlation planes of Eigenvector 1. Solid
symbols are radio-quiet sources (boxes: Pop. A, circles: Pop. B).
Open symbols are radio-loud sources (open: core dominated,
crossed: lobe dominated). Panel d: the BG92 sample plotted in the
W(\feiiq) vs. FWHM(\hbbc) plane. The dotted line at 2000 \kms\
separates NLSy1 and the rest of the sample, while the dot-dashed
line at 4000 \kms\ sets the limit between Pop. A and Pop. B.
Error bars indicate 2$\sigma$\ uncertainties for a ``typical'',
non-pathological data point at FWHM(\hbbc) $\approx$ 4000 \kms,
\rfe $\approx $0.5, and error on \gs\ equal to the median of
errors for which \gs\ was available. \label{fig:e1}}

\end{figure}

Figures \ref{fig:e1} a,b,c show the 2D projections of the E1
correlation space.  AGN plotted in the figures come from three
data sources:  1) the BG92 sample of 87 mostly radio-quiet (RQ)
AGN (16 radio-loud: RL), 2) the MS96  mixed sample of 21 RQ and
31 RL sources as well as 3) a sample of 24 new sources with
matching HST UV archival and optical ground-based spectra (8 RL).
The data sources combine the most complete sample of AGN with
high quality spectral data (BG92) and two overlapping samples of
comparable quality data with matching optical \hb\ and UV \civ.
Soft X-ray photon indices are available for a large fraction of
the above samples (Wang et al. 1996; Brinkmann et al.  1997;
Siebert et al. 1998; Yuan et al.  1998). Values determined with
hydrogen column density N$_H$\ as a free parameter were preferred
when available. The hard X-ray photon index provides an
alternative, but less sensitive, E1 correlate (Piccinotti et al.
1982; Brandt et al. 1997). Our total sample includes 128 sources
(45 RL) with optical measures, of which, 76 (39 RL) have UV
measures.   Mean error bars (2$\sigma$) are indicated with:  i)
average errors for sources in the middle of the diagrams for
\feiiq\ and \hb\ measures as well as ii) a median value for all
\gs\ measures. See BG92 and MS96 for detailed discussion of
reduction and analysis procedures.

RQ sources are indicated by solid symbols in the figures while
open and crossed circles indicate flat (core-dominated) and steep
(lobe-dominated) spectrum RL (as defined by Kellermann et al 1989)
sources respectively. The majority of sources show a well defined
correlation in Figures \ref{fig:e1} a,b,c. One can make a first
order interpretation of E1 correlations in terms of RL vs. RQ
differences. Sources that fall in the large FWHM \hb\ and large
\rfe\ region (like PG 0043+039; Turnshek et al 1994) are
apparently very rare and pathological (possibly associated to
mixed Starburst/AGN properties). Fig.  \ref{fig:e1} 1 a,b,c
indicate that these two populations show a clear separation with
RQ having, for example, a mean FWHM(\hbbc) $\approx$ 2300 \kms\
less than RL sources (a few RL sources with FWHM(\hbbc) $\sim$
11--20000 \kms\ are not shown in the figures). An alternative
interpretation of E1 sees two populations of RQ AGN:  1)
population A (filled boxes) which shows little overlap with the RL
domain (65\%\ of the BG92 RQ sample) and 2) population B (filled
circles) that occupies essentially the same E1 domain as the RL
sources (~25\%\ of the BG92 RQ sources). Fig. \ref{fig:e1} 1 d
shows a plot of W(\feiiq) vs.  FWHM(\hbbc) for the BG92 sample
that provided the motivation for the population A--B concept.
Rather than a correlation this plot shows two disjoint
populations of sources with no correlation within either
population. Correlations appear when \rfe\ is used instead of
W(\feii) or W(\hbbc).  We introduce the population A and B
distinction for the sake of argument but suggest that it may have
more fundamental significance.  Table 1 presents mean parameter
values (and associated sample standard deviations) in arbitrary
bins of FWHM(\hbbc) that allow one to compare RL vs. RQ as well
as Pop. A vs. Pop. B sources.

It is not easy at this time to determine what role sample
selection biases play in E1 but it is clear that BG92 selection
techniques favored RQ sources with narrow Balmer line profiles.
Recent work indicates that selection on the basis of soft X-ray
excess favors them even more strongly (e.g. Moran et al.  1996;
Grupe et al. 1999).  At the same time core-dominated/flat
spectrum RL sources are no doubt over-represented if their
emission is beamed. This study makes no claim to completeness.
Beyond BG92 our sample selection is driven by the availability of
i) high s/n and resolution ground based data and ii) matching HST
UV spectra.  Table 2 presents the results of a correlation
analysis for the three principal E1 and related parameters (the
Pearson's correlation coefficient r$_P$\ is reported along with
the probability P of a chance correlation in a sample of N
objects). As the figures and Table 2 also suggest, the strongest
E1 correlations are found for the RQ sources. The highest ranked
E1 correlation (\gs\ vs. FWHM \hbbc) is present only in the RQ
sample which is not surprising since RL sources show no evidence
for a soft X-ray excess. The nearly identical r$_P$\ for the ALL
RQ and Pop. A RQ samples suggest that the principal correlations
are driven by Pop. A. RQ Pop.  B and RL sources show the same E1
parameter domain and neither sample shows evidence for
significant internal correlations. The latter result is due, at
least in part, to: a) the absence of a soft X-ray excess in these
sources and b) the weakness of \feii\ emission in these
relatively broad-lined sources. If the minimum detectable \feii\
emission for sources with profile widths FWHM(\feiiq) $\approx$
FWHM(\hbbc) $\approx$1000 \kms is W(\feiiq) $\approx$10--15\AA\
then we estimate that it will be $\approx$ 25 and 40 \AA\ for
FWHM(\hbbc) $\approx$ 5000 and 10000 \kms, respectively.  The
message here is that values of \rfe$\simlt$0.2 are very uncertain
even with the best available data.

There are several lines of evidence that suggest continuum
luminosity is uncorrelated with E1 parameter space. 1) Optical
luminosity appears in the second orthogonal eigenvector
identified in BG92, 2) BAL sources occupy a similar domain to the
NLSy1 but they are ``displaced'' in:  a) X-ray luminosity - BALs
are X-ray quiet and b) optical luminosity where they are on
average about 4$\times$\ more luminous than the NLSy1 sources in
BG92. 3) RL and Pop. B RQ sources occupy the same parameter
domain but the RL sources show $\sim 5\times$\ higher optical
continuum luminosities which may in part reflect the presence of
beamed sources in the RL sample. 4) We also find no evidence for
a correlation between E1 coordinates and radio continuum
luminosity for RQ sources (see Kellermann et al.  1989; Falcke et
al. 1996). This last result tells us that RL sources are a
distinct AGN population that shows fundamental differences in BLR
structure and kinematics. E1 indicates that the weak radio
emission from RQ sources is unrelated to the RL phenomenon.

E1 also shows evidence for a parameter space separation between
steep and flat spectrum RL sources. This shows up in Table 2 as a
possible RL correlation between FWHM \hbbc\ and \rfe. Steep
spectrum RL sources represent the opposite extremum from NLSy1
while flat spectrum sources are more similar to RQ sources.  The
tendency for steep spectrum RL sources to show the broadest
Balmer profiles and the weakest \feii\ emission is confirmed in
two large (with some source overlap between themselves and our RL
sample) surveys (Brotherton 1996; Corbin 1997). Siebert et al
(1998) provide evidence that this separation may also be present
in the X-ray spectral index. NLSy1 are a RQ and population A
extremum with the narrowest Balmer line profiles, strong \feii\
emission and a strong soft X-ray excess.  NLSy1 show a clear
continuity and correlation with broader line Seyfert 1 galaxies
in all measured parameters. This challenges the idea that they
represent a unique or disjoint AGN population.  BAL quasars (4 RQ
BAL in the BG92 sample) occupy an E1 domain that is similar to
the NLSy1.  Other (BAL) studies (e.g. Boroson \& Myers 1992) have
also suggested that BAL quasars show Balmer line FWHM$ \simlt$
3000 \kms and moderate to strong \rfe measures.

\begin{figure}
\figurenum{2} \plotone{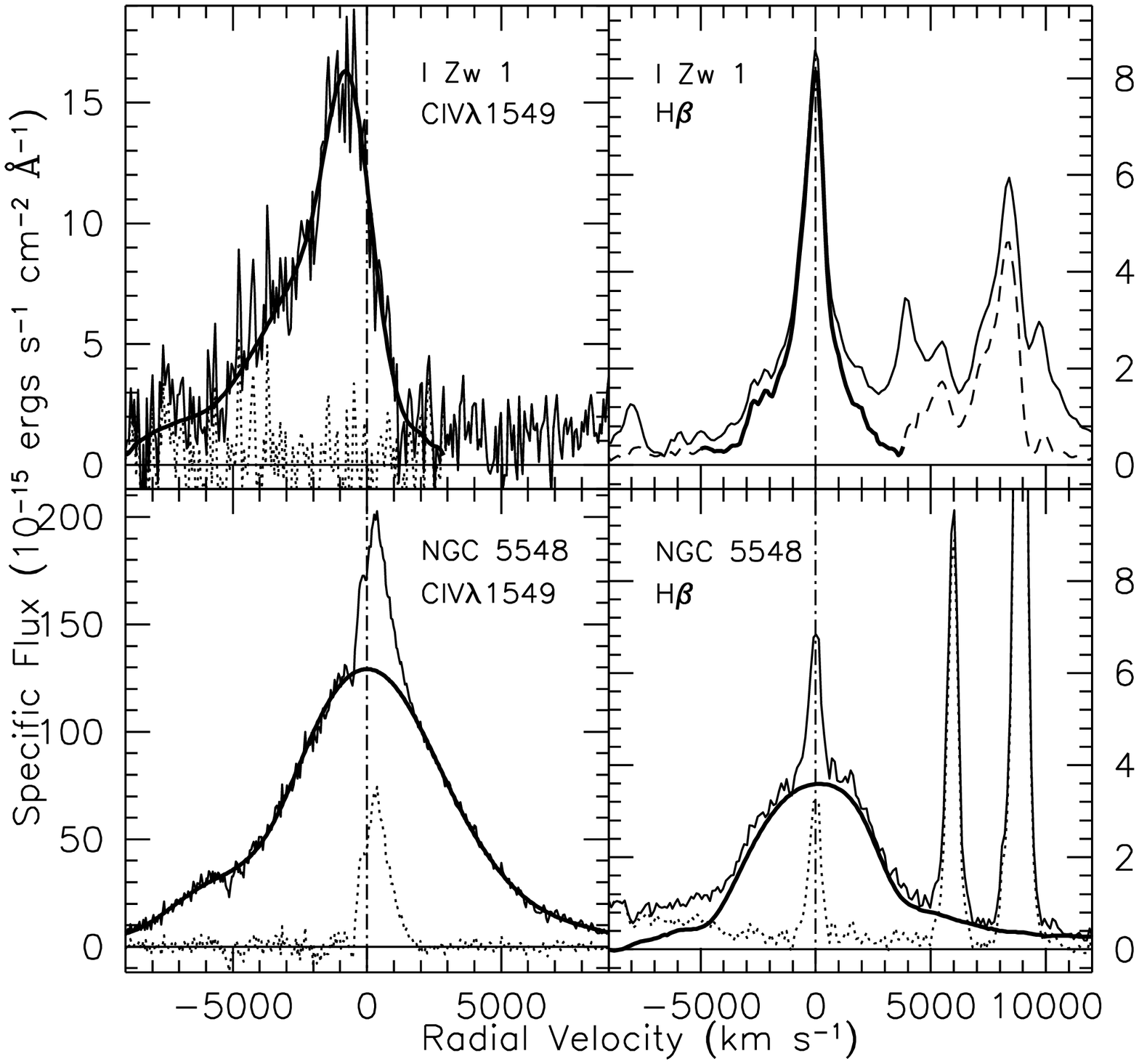} \caption[2]{\hb\ (right
panels; solid line: \hbbc) and \civ\ (left panels) broad line
profiles for NLSy1 source IZw1 (``extreme'' Pop. A, upper half)
and Pop. B Seyfert 1 source NGC5548 (lower half).  Note that the
abscissa origin of I Zw 1 has been set by {\sc Hi} 21 cm
measures. \oiii\ lines show an appreciable blueshift with respect
to the {\sc Hi} reference radial velocity as well as  to the peak
of the \hb\ profile. Dotted lines: residuals after broad
component (thick line subtraction). Dashed line in I Zw 1 \hb\
panel: residual spectrum after \feii\ subtraction.

\label{fig:profiles}}
\end{figure}

\begin{figure}
\figurenum{3} \plotone{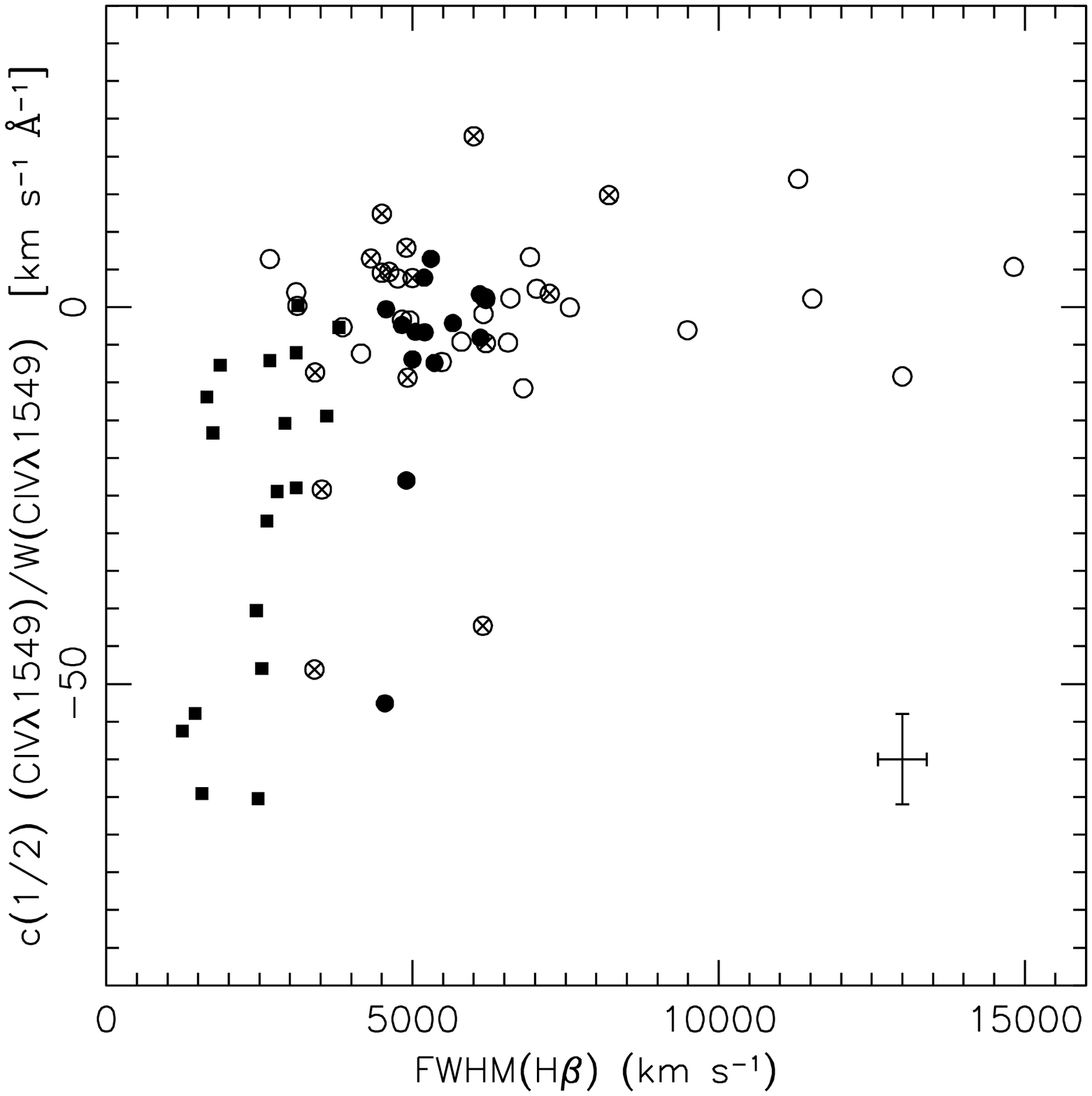} \caption[3]{Relationships
between FWHM(\hbbc) and \civ\ FWHM profile shift (blueshifts are
negative) normalized by the rest-frame equivalent width of \civ,
W(\civ). Error bars are indicative as in Fig. \ref{fig:e1} and
refer to W(\civ) $\approx$ 75 \AA\ and shift $\approx $ 400 \kms.
\label{fig:civ}}
\end{figure}

Much less clear is whether RQ population B sources represent a
disjoint AGN population or show a smooth continuation of the
population A correlations. Figure \ref{fig:e1} b for example
shows a breakdown of the \gs\ -- FWHM(\hb) correlation at the
nominal boundary between population A and B. Whatever the
relation between the two RQ populations, their line profiles show
striking differences.  Figure \ref{fig:profiles} shows a
comparison of the \hbbc\ and \civbc\ line profiles for prototype
NLSy1 source I Zw 1 and NGC 5548 which is a typical broad line
Seyfert galaxy.  While differences in the Balmer profiles are
striking, the most impressive difference is the apparent
kinematic {\it decoupling} of the \civ\ and \hb\ profiles in I Zw
1. Our previous work (MS96) suggested that properly NLR corrected
\civ\ in RQ sources (see Sulentic \& Marziani 1999) is always
blueshifted relative to \hb\ and the AGN rest frame. Fig.
\ref{fig:civ}    shows the \civbc\ line shift vs. FWHM(\hbbc)
diagram. We have normalized the \civ\ shift by W(\civ) in this
plot because we see a complementary trend in E1 for W(\civ) to be
smallest in the NLSy1 population. Figure \ref{fig:civ} confirms
that essentially all RQ sources show a \civ\ blueshift while RL
sources show equal red and blueshifts with amplitudes generally
less than $\pm$10$^3$\ \kms. Correlations of \civ\ shift with
\rfe\ and \gs\ (see Table 2) indicate that it is likely to be an
important E1 correlate. We will explore it further in succeeding
papers.

\section{From Observational to Physical Parameters}

E1 correlates AGN spectroscopic data in a way that removes much
of the apparent ``randomness'' of line properties. It also
redefines input parameters for photoionization and kinematical
models. We have only begun to explore the E1 parameter space and
and the physics that drives it.  Accumulating evidence from
numerous studies suggests that the correlations in E1 involve at
least two principal independent parameters: the AGN luminosity to
black hole mass ratio and the source orientation. The role of Fe
abundance, disk magnetic fields and black hole angular momentum
are beyond the scope of this introduction to E1.

Available evidence suggests that the central source luminosity to
black hole mass ratio {L/M} may be systematically higher in Pop.
A sources (i.e. a higher accretion rate). A role for L/M is
supported by:  a) the existence of a soft X-ray excess,
particularly in NLSy1, that has been related to a higher
accretion rate (Pounds et al. 1995) and b) the thermal signature
of the disk itself (Page et al. 1999; Puchnarewicz et al. 1998).
If the soft X-ray excess is the high energy end of a thermal
signature of the accretion disk then a steep rising blue optical
continuum (in low z sources) can be interpreted as the low energy
end of the ``big blue bump''.  This would explain why BG92, with
quasars selected on the basis of U--B colors, would favor
detection of NLSy1. \civ\ shows the largest blueshifts in NLSy1
sources, and the lowest \civ\ W values, both consistent with the
idea that the line emission arises in a disk wind (e.g. Murray \&
Chiang 1997; Bottorff et al. 1997). The observations are
consistent with the idea that L/M is only high enough in RQ
sources (especially Pop. A) to trigger a radiation pressure
driven outflow. The Balmer profiles in Pop. A sources are most
easily interpreted as arising in an optically thick illuminated
accretion disk (see e.g.  Sulentic et al.  1998).  RL sources show
Balmer lines with {\it both} large red and blue profile shifts.

RL sources show no evidence for a soft X-ray excess which may
reflect a real absence of this component or that it is dominated
by a relatively hard component related to X-ray emission related
to the radio loudness.  We favor the former interpretation
because most RQ population B sources also show a soft X-ray
deficit. The absence of a soft X-ray component in RL+RQ pop B
sources coupled with the absence of the CIV blueshift are
consistent with weak or absent disk structure. The broad lines
may arise in a biconical structure in many or all of these
sources (Marziani et al. 1993; Sulentic et al. 1995). Evidence
for a bicone origin of the BLR includes: a) sources with double
peaked broad lines (e.g. Sulentic et al. 1995; Eracleous \&
Halpern 1998), b) sources with single peaked broad lines showing
both large red and blue displacements (e.g. Marziani et al. 1983;
Halpern et al. 1998), c) sources with a transient BLR (e.g.
Storchi-Bergmann et al. 1993; Ulrich 2000) and spectropolarimetry
of a) and b) sources (Corbett et al. 1998).

The E1 parameter space is measuring aspects of both the geometry
and kinematic of the broad line region in AGN. The strength of the
correlations and  reasonable orthogonality of the parameters
suggest that a better diagnostic space for AGN is unlikely to be
found. Even in the preliminary presentation, E1 allows us to
resolve some AGN conundrums. 1) RL sources are found to be an AGN
population with fundamental geometrical and kinematic difference
from the RQ majority. 2) NLSy1 are found to be an extremum of
the RQ population rather than a pathological or disjoint AGN
population.  3) If the width of the Balmer lines bear any
signature of source orientation then some NLSy1 (at least, I Zw
1) are seen at or near pole-on (face-on accretion disk)
orientation, while some BAL QSOs are likely to be misaligned
NLSy1. 4) Population B RQ quasars with Balmer lines broader than
$\approx$ 4000 \kms  may represent a distinct class (RL pre/post
cursors) from narrower population A RQ sources. 5) \civ\ line
shifts indicate the presence in Pop. A (and possibly Pop. B) RQ
sources of an ubiquitous disk outflow/wind.

\acknowledgements

PM acknowledges financial support from the Italian Ministry of
University and Scientific and Technological Research (MURST)
through grant Cofin 98-02-32 and from the IA-UNAM where part of
this work was done.

\newpage
\section{References}

\REF Boller, T., Brandt, W.N. \& Fink, H. 1996, \aap, 305, 53

\REF Boroson, T.A. \&\ Green, R.F. 1992, ApJS, 80, 109

\REF Boroson, T.A. \& Myers, K. 1992, ApJ, 397, 442

\REF Bottorff, M., et al. 1997, ApJ, 479, 200

\REF Brandt W.N., Mathur S. \& Elvis M. 1997, MNRAS, 285, L25

\REF Brinkmann, W., Yuan, W, \& Siebert, J. 1997, A\&Ap, 319, 413

\REF Brotherton, M. S. 1996, ApJS, 102, 1

\REF Corbin, M. 1997, ApJS, 113, 245

\REF Eracleous, M. \& Halpern, J. 1998, ApJ, 505, 577

\REF Falcke, H., Sherwood, W. \& Patnail, A. 1996, ApJ, 471, 106

\REF Grupe, D., et al. 1999, A\&Ap, 350, 805

\REF Halpern, J., Eracleous, M. \& Forster, K. 1998, ApL. 501, 103

\REF Kellermann, K. et al. 1989, AJ, 98, 1195

\REF Korista, K.T. et al. 1995, ApJS, 97, 285

\REF Laor, A. et al. 1997, ApJ, 477, 93

\REF Marziani, P., et al. 1993, 410, 56

\REF Marziani, P., Sulentic, J.W., Dultzin-Hacyan, D., Calvani,
M. \& Moles, M. 1996, ApJS, 104, 37

\REF Moran, E., Halpern, J. \& Helfland, D. 1996, ApJS, 106, 341

\REF Murray, N. \& Chiang, J. 1997, 474, 91

\REF Page, M.J. et al. 1999, MNRAS, 305, 775

\REF Piccinotti, G., Mushotzky, R.F., Boldt, E.A., Holt, S.S.,
Marshall, F.E., Serlemitsos, P.J. \& Shafer, R.A. 1982, Ap.J,
253, 485

\REF Pounds, K.A., Done, C. \& Osborne, J.P. 1995, MNRAS, 277, L5

\REF Puchnarewicz, E.M. et al. 1998, MNRAS, 293, L52

\REF Siebert, J., et al. 1998, A\&Ap, 301, 261

\REF Storchi-Bergmann, T., Baldwin, J. \& Wilson, A. 1993, ApJ,
410, L11

\REF Sulentic, J.W. \& Marziani, P. 1999, ApJ, 518, L9

\REF Sulentic, J.W., et al. 1995, ApJ, 438, L1

\REF Sulentic, J.W., et al. 1998, ApJ, 501, 54

\REF Turnshek, D. et al. 1994, ApJ, 428, 93

\REF Ulrich, M-H. 2000, A\&ApRev, in press

\REF Wang, T., Brinkmann, W., Bergeron, J. 1996, A\&Ap, 309, 81

\REF Yuan, W., Siebert, J. \& Brinkmann, W. 1998, A\&Ap, 334, 498

\newpage

\end{document}